\documentclass[twocolumn]{aastex631}

\usepackage{CJKutf8}

\usepackage{amsmath}
\usepackage{multirow}

\newcommand{\ftm}{F2M1106}

\newcommand{\paa}{{Pa$\alpha$}}
\newcommand{\pab}{{Pa$\beta$}}

\newcommand{\av}{{A$_{\rm V}$}}
\newcommand{\ebv}{E(B$-$V)}

\newcommand{\siv}{[\ion{S}{4}] $\lambda$10.51 $\mu$m}
\newcommand{\oiiitext}{[\ion{O}{3}]}
\newcommand{\sivtext}{[\ion{S}{4}]}

\newcommand{\mum}{\ifmmode{\rm \mu m}\else{$\mu$m}\fi}
\newcommand{\vdisp}{$\vdisp$}
\newcommand{\wba}{W$_{80}$}
\newcommand{\vwu}{{v$_{50}$}}

\newcommand{\fsb}{erg s$^{-1}$ cm$^{-2}$ }

\newcommand{\oiii}{{[\ion{O}{3}] $\lambda$5007}}

\newcommand{\hb}{\hbox{H$\beta$}}

\newcommand{\feiitext}{\hbox{[Fe$\,${\scriptsize II}]}}

\newcommand{\kms}{km\,s$^{-1}$} 
\newcommand{\msun}{M$_{\odot}$} 
\newcommand{\msunyr}{{M$_{\sun}$ yr$^{-1}$}}

\newcommand\jwst{\emph{JWST}}

\newcommand\qtdfit{\texttt{q3dfit}}

\shorttitle{F2M1106}
\shortauthors{Liu al.}
\graphicspath{{./}{figures/}}

\begin{document}

\title{First results from the JWST Early Release Science Program Q3D: The Fast Outflow in a Red Quasar at z$=$0.44}

\author[0000-0003-3762-7344]{Weizhe Liu \begin{CJK}{UTF8}{gbsn}(刘伟哲)\end{CJK}}
\affiliation{Steward Observatory, University of Arizona, 933 N. Cherry Ave., Tucson, AZ 85721, USA}

\correspondingauthor{Weizhe Liu}
\email{wzliu@arizona.edu}

\author[0000-0002-3158-6820]{Sylvain Veilleux}
\affiliation{Department of Astronomy and Joint Space-Science Institute, University of Maryland, College Park, MD 20742, USA}

\author[0000-0002-4419-8325]{Swetha Sankar}
\affiliation{Department of Physics and Astronomy, Bloomberg Center, Johns Hopkins University, Baltimore, MD 21218, USA}

\author[0000-0002-1608-7564]{David S. N. Rupke}
\affiliation{Department of Physics, Rhodes College, Memphis, TN 38112, USA}

\author[0000-0001-6100-6869]{Nadia L. Zakamska}
\affiliation{Department of Physics and Astronomy, Bloomberg Center, Johns Hopkins University, Baltimore, MD 21218, USA}
\affiliation{Institute for Advanced Study, Princeton, NJ 08540, USA}

\author[0000-0003-2212-6045]{Dominika Wylezalek}
\affiliation{Zentrum für Astronomie der Universität Heidelberg, Astronomisches Rechen-Institut, Mönchhofstr 12-14, D-69120 Heidelberg, Germany}

\author[0000-0002-0710-3729]{Andrey Vayner}
\affiliation{IPAC, California Institute of Technology, 1200 E. California Blvd., Pasadena, CA 91125, USA}

\author[0000-0002-6948-1485]{Caroline Bertemes}
\affiliation{Zentrum für Astronomie der Universität Heidelberg, Astronomisches Rechen-Institut, Mönchhofstr 12-14, D-69120 Heidelberg, Germany}

\author[0000-0002-9932-1298]{Yu-Ching Chen}
\affiliation{Department of Physics and Astronomy, Bloomberg Center, Johns Hopkins University, Baltimore, MD 21218, USA}

\author[0000-0001-7572-5231]{Yuzo Ishikawa}
\affiliation{Department of Physics and Astronomy, Bloomberg Center, Johns Hopkins University, Baltimore, MD 21218, USA}


\author[0000-0002-5612-3427]{Jenny E. Greene}
\affiliation{Department of Astrophysical Sciences, Princeton University, 4 Ivy Lane, Princeton, NJ 08544, USA}

\author[0000-0001-8813-4182]{Timothy Heckman}
\affiliation{Department of Physics and Astronomy, Bloomberg Center, Johns Hopkins University, Baltimore, MD 21218, USA}


\author{Guilin Liu}
\affiliation{CAS Key Laboratory for Research in Galaxies and Cosmology, Department of Astronomy, University of Science and Technology of China, Hefei, Anhui 230026, China}
\affiliation{School of Astronomy and Space Science, University of Science and Technology of China, Hefei 230026, China}

\author[0000-0001-8813-4182]{Hsiao-Wen Chen}
\affiliation{Department of Astronomy \& Astrophysics, The University of Chicago, 5640 South Ellis Avenue, Chicago, IL 60637, USA}

\author[0000-0003-0291-9582]{Dieter Lutz}
\affiliation{Max-Planck-Institut für Extraterrestrische Physik, Giessenbachstrasse 1, D-85748 Garching, Germany}

\author[0000-0001-9487-8583]{Sean D. Johnson}
\affiliation{Department of Astronomy, University of Michigan, Ann Arbor, MI 48109, USA}

\author[0000-0001-5783-6544]{Nicole P. H. Nesvadba}
\affiliation{Université de la Côte d'Azur, Observatoire de la Côte d'Azur, CNRS, Laboratoire Lagrange, Bd de l'Observatoire, CS 34229, Nice cedex 4 F-06304, France}

\author[0000-0002-3471-981X]{Patrick Ogle}
\affiliation{Space Telescope Science Institute, 3700, San Martin Drive, Baltimore, MD 21218, USA}

\author[0009-0003-5128-2159]{Nadiia Diachenko}
\affiliation{Department of Physics and Astronomy, Bloomberg Center, Johns Hopkins University, Baltimore, MD 21218, USA}

\author[0000-0003-4700-663X]{Andy D. Goulding}
\affiliation{Department of Astrophysical Sciences, Princeton University, 4 Ivy Lane, Princeton, NJ 08544, USA}

\author[0000-0003-4565-8239]{Kevin N. Hainline}
\affiliation{Steward Observatory, University of Arizona, 933 North Cherry Avenue, Tucson, AZ 85721, USA}

\author{Fred Hamann}
\affiliation{Department of Physics \& Astronomy, University of California, Riverside, CA 92521, USA}

\author{Hui Xian Grace Lim}
\affiliation{Department of Physics, Rhodes College, Memphis, TN, 38112, USA}

\author[0000-0001-6126-5238]{Nora Lützgendorf}
\affiliation{European Space Agency, Space Telescope Science Institute, Baltimore, Maryland, USA}

\author[0000-0002-1047-9583]{Vincenzo Mainieri}
\affiliation{European Southern Observatory, Karl-Schwarzschild-Straße 2, D-85748 Garching bei München, Germany}

\author{Ryan McCrory}
\affiliation{Department of Physics, Rhodes College, Memphis, TN, 38112, USA}

\author{Grey Murphree}
\affiliation{Department of Physics, Rhodes College, Memphis, TN, 38112, USA}
\affiliation{Institute for Astronomy, University of Hawai'i, Honolulu, HI, 96822, USA}

\author[0000-0002-0018-3666]{Eckhard Sturm}
\affiliation{Max-Planck-Institut für Extraterrestrische Physik, Giessenbachstrasse 1, D-85748 Garching, Germany}

\author{Lillian Whitesell}
\affiliation{Department of Physics, Rhodes College, Memphis, TN, 38112, USA}

\begin{abstract}

Quasar feedback may play a key role in the evolution of massive galaxies. The dust-reddened quasar, F2M110648.35$+$480712 at $z = 0.4352$ is one of the few cases at its redshift that exhibits powerful quasar feedback through bipolar outflows. Our new observation with the integral field unit mode of Near-infrared Spectrograph onboard JWST opens a new window to examine this spectacular outflow through Pa$\alpha$ emission line with $\sim$3$\times$ better spatial resolution than previous work. The morphology and kinematics of the Pa$\alpha$ nebula confirm the existence of a bipolar outflow extending on a scale of $\sim$17$\times$14 kpc and with a velocity reaching $\sim$1100 km s$^{-1}$. The higher spatial resolution of our new observation leads to more reliable measurements of outflow kinematics. Considering only the spatially resolved outflow and assuming an electron density of 100 cm$^{-2}$, the mass, momentum and kinetic energy outflow rates are $\sim$50--210 M$_{\sun}$ yr$^{-1}$, $\sim$0.3--1.7$\times$10$^{36}$ dynes ($\sim$14--78\% of the quasar photon momentum flux) and $\sim$0.16--1.27$\times$10$^{44}$ erg s$^{-1}$ ($\sim$0.02--0.20\% of the quasar bolometric luminosity), respectively. The local instantaneous outflow rates generally decrease radially. We infer that the quasar is powerful enough to drive the outflow, while stellar processes cannot be overlooked as a contributing energy source. The mass outflow rate is $\sim$0.4--1.5 times the star formation rate, and the ratio of kinetic energy outflow rate to the quasar bolometric luminosity is comparable to the minimum value required for negative quasar feedback in simulations. This outflow may help regulate the star formation activity within the system to some extent.

\end{abstract}



\section{Introduction} 
\label{sec:intro}

Quasar feedback is believed to play a critical role in the evolution of massive galaxies. Such feedback may quench/regulate star formation activity in galaxies \citep[e.g.][]{DiMatteo2005}, shape the circum/intergalactic medium \citep[CGM/IGM; e.g.][]{Tumlinson2017}, and regulate supermassive black hole (SMBH) accretion \citep[e.g.][]{Hopkins2016}. Powerful, quasar-driven outflows are deemed one of the most effective of such quasar feedback \citep[e.g.,][]{Vei2005,Fab2012,Harrison2018,Veilleux20}. A comprehensive understanding of the physical properties of such outflows are crucial to evaluate their impact on galaxy evolution.

The luminous \oiii\ emission line in the rest-frame optical is one of the most popular tracers of quasar/AGN-driven outflows \citep[e.g.][and references therein]{Harrison2014,Liu2013b, nadiajenny2014,Carniani2015, Cresci2015,Wylezalek16a, Zakamska2016, Bischetti2017,Vayner2017,Rupke2017,Perrotta2019,Finnerty2020, Wylezalek2020,Liu2020,Vayner2021b,Vayner2021c,Wangwuji2024}.
Despite the wide usage, the outflow properties (i.e., morphology, dynamics and energetics) based on \oiii\ usually suffer from two uncertainties. First, the metallicity of the ionized outflowing gas is uncertain, leading to difficulty in recovering the total outflowing gas mass. Second, many powerful quasar-driven outflows are likely dusty or are launched in dusty environments \citep[e.g.,][]{fauc12a,fauc12b}, which may severely attenuate \oiii\ emission line in the optical, making the detection of faint emission from the outflow even more challenging. Given these two major caveats, alternative tracers are needed for a comprehensive view of these outflows.

As another important tracer of galactic outflows, \paa\ in the near-infrared has been proven successful in deciphering quasar-driven outflows in the nearby universe \citep[e.g.][and references therein]{Santoro2018,Nardini2018,Bohn2021,Bianchin2024}. In the low-z universe, compared to \oiii\ in the optical usually without AO observations, \paa\ in the near-infrared allows for better spatial resolution with ground-based AO observations or space-based observations now in the JWST era, is insensitive to the gaseous phase metallicity when tracing gas mass, and suffers much less from dust extinction. 

Despite the advantages stated above, adopting \paa\ to carry out spatially resolved studies of quasar/AGN-driven outflows has been confined to the nearby universe where the line is bright enough and falls in the wavelength coverage of appropriate instruments (e.g. ground-based near-IR integral field spectrograph or IFS in short).
With the advent of IFS onboard JWST, it finally becomes much easier to unveil the properties of quasar-driven outflows adopting \paa\ beyond the local universe.

``Q-3D: Imaging Spectroscopy of Quasar Hosts with JWST" is an Early Release Science Program (PID 1335, PI Wylezalek, co-PIs Veilleux, Zakamska; Software Lead: Rupke), on \jwst\ \citep{Wylezalek2022} to establish infrared diagnostics of quasar-driven outflows and to develop techniques for probing their impact on the quasar host evolution. One of the program's primary goals is to probe quasar-driven outflows at high spatial and spectral resolution using near-IR capabilities of \jwst, adopting tracers including \paa\ emission.

Specifically, F2M110648.35$+$480712.3 (or \ftm\ hereafter) is the nearest quasar in the Q3D sample with $z=0.4352$. {The redshift is determined from the stellar kinematics based on Keck/KCWI data (Rupke et al., in preparation).} It has a bolometric luminosity of log($L_\mathrm{bol}/$erg~s$^{-1}$) $= 46.8$ \citep[based on rest-frame 12 \mum\ luminosity;][]{Shen2023}. With $J - K = 2.31$ mag, $R - K = 4.01$ mag, and $\ebv = 0.44$ mag, it belongs to the red quasar population \citep{Glikman2012}. A fast, powerful outflow is clearly detected in multiple rest-frame optical and infrared emission lines including \oiii\ \citep{Shen2023}, \siv\ \citep{Rupke2023}, and \feiitext\ (Sankar et al. in prep.), extending out to distances over 10~kpc from the nucleus. 
\ftm\ is an ideal laboratory to examine in detail the outflow properties through \paa\ emission line at intermediate redshift.

The main objectives of this paper on the \jwst\ data of \ftm\ are to characterize the warm ionized phase of the outflow in this system with \paa\ emission and assess the impact of the outflow on the host galaxy. This paper is organized as follows. In Section \ref{sec:obs_redux}, we briefly describe the observations and steps taken to reduce the \jwst\ data. In Section \ref{sec:analysis}, we outline our data analysis flow with the software package \qtdfit\ \citep{q3dfit}. The results from this analysis are presented in Section \ref{sec:results}, and the discussion on \paa-based outflow properties and its impact on the host galaxy is then presented in Section \ref{sec:discussion}. The conclusions are summarized in Section \ref{sec:conclusions}.  

Throughout this paper, we assume the $\Lambda$CDM cosmology with $H_{0} = 70$~km~s$^{-1}$~Mpc$^{-1}$, $\Omega_m = 0.3$ and $\Omega_{\lambda} = 0.7$. The resulting physical scale is 1\arcsec\ $=$ 5.649 kpc.

\section{Observations and Data Reduction} 
\label{sec:obs_redux}

\subsection{Observation}
\label{subsec:observations}

\ftm\ was observed on 2022-11-13 by \jwst\ using the NIRSpec Instrument in IFS mode \citep{Bok2022, Jak2022}. These data are publicly available on the Mikulski Archive for Space Telescopes (MAST) at the Space Telescope Science Institute, which can be accessed via \dataset[10.17909/b6v0-8f22]{http://dx.doi.org/10.17909/b6v0-8f22}. The field of view (FOV) of the IFS observations are $\sim 3\arcsec \times 3 \arcsec$ or $\sim 17 \times 17$~kpc for this object. The filter/grating configurations of the observations are G235H/F170LP G395H/290LP, with corresponding wavelength coverage of $1.66-5.14~\mu$m or $1.16-3.58~\mu$m at the redshift of \ftm. The grating has a nominal resolving power $\lambda / \Delta\lambda \simeq 2700$, corresponding to a velocity resolution $\sim 110$ \kms, which allows us to well spectrally resolve the emission lines with typical velocity widths of several hundred \kms. For each configuration, we use a 9-point small cycling dither pattern with 25 groups and 1 integration per position to improve the spatial sampling and enable a better characterization of the point spread function (PSF). One leakage exposure at the first dither position is also taken to account for light leaking through the closed micro-shutter array (MSA) and from the failed open shutters. The total integration time per configuration is 2101 seconds on source. 

\subsection{Data Reduction}
\label{subsec:data_reduction}

The NIRSpec IFS data of \ftm\ are reduced following the same method adopted for the other two targets from the Q3D program \citep[][]{Vay2023a,VeilleuxLiu2023}. 
In brief, we reduce the data with \jwst\ Calibration pipeline version 1.9.62 in conjunction with the
jwst 1063.pmap version of the calibration reference files. In addition to the processes in the public pipeline, we take extra/modified steps to generate the final data properly. Specifically, in step \verb|Spec1Pipeline|, we perform standard infrared detector reduction steps such as dark current subtraction, fitting ramps of non-destructive group readouts, combining groups and integrations, data quality flagging, cosmic ray removal, bias subtraction, linearity, and persistence correction. We further apply a band correction to mitigate significant instrument-induced
variations. Then in step \verb|Spec2Pipeline|, we assign a world coordinate system to each frame, apply flat field correction, flux calibration, and extract the 2D spectra into a 3D data cube using the \verb|cube build| routine. 
Finally, we use the \verb|reproject| package\footnote{\href{https://reproject.readthedocs.io/en/stable/}{https:reproject.readthedocs.io/en/stable/}} to combine the different dithers into a single master data cube with a spaxel (i.e., spatial pixel) size of 0\farcs05. using their drizzle algorithm. In the process, the outliers were further removed by sigma-clipping across different dithers.

\section{PSF Subtraction and Spectral Fitting}
\label{sec:analysis}

To reveal the faint, extended emission outside the quasar, we first carefully remove the bright quasar light from the NIRSpec data cube adopting \qtdfit\ \citep{q3dfit}, a software package designed for the removal of bright point spread function (PSF) from \jwst\ data cubes.  
We analyze the NIRSpec data of \ftm\ with \qtdfit\ following the same method adopted for the other two objects in the Q3D program, J1652  \citep[][]{Wylezalek2022,Vay2023a,Vay2023b} and XID~2028 \citep{VeilleuxLiu2023}. Here we summarize the key aspects of \qtdfit\ briefly.

First, we build a quasar template spectrum with a r$=$4.5 pixels (0\farcs225) aperture centered on the brightest spaxel, which is used as the PSF spectral model. 
Single-spaxel spectra near the quasar in NIRSpec data suffer from oscillations as a function of wavelength, which are caused by the spatial undersampling of the PSF by the native 0\arcsec.1 pixel size of the IFS detector \citep[][]{Law2023,Perna2023}.
As a result, for the quasar template spectrum, we choose the smallest aperture size that suppresses the oscillating spectral pattern to an acceptable extent. {The quasar template spectrum is built directly from the data cube, which leads to the apparent mild “noise” in the fitting model. However, these small fluctuations are much smaller than the typical noise level of spectra from individual spaxels (Fig. \ref{fig:fig2}) and thus do not affect the fitting results. In addition, we have also tested fits with a smoothed quasar template spectrum, and all of our results remain unchanged.}

Next, for each spaxel within the data cube, we fit the corresponding spectrum with (i) a scaled quasar template spectrum representing the PSF contribution, (ii) a set of emission lines with Gaussian profiles from the host galaxy, and (iii) featureless monotonic polynomials representing the stellar continuum contribution. Specifically, here the quasar template spectrum is scaled with multiplicative polynomials and exponential functions to account for variations in the spectral shape of point source emission in individual spaxels due to the wavelength dependence of the NIRSpec PSF. 

For the host emission lines, we carry out two separate fits. In the first run, we only fit the \paa\ line, the brightest one in the NIRSpec data cube, and mask all other emission lines to maximize the detection of faint extended line emission. We set the velocity dispersions of the lines to be 60 -- 1000 \kms\ to avoid fitting narrow noise spikes and very broad artificial features caused by the undersampling of PSF as mentioned earlier. The number of Gaussian components is selected to minimize the final reduced chi-square, and we only keep a Gaussian component with a peak flux density above 3$\sigma$. We visually inspect individual fits to further reject erroneous ones. {Neither line kinematics nor line fluxes change by more than a few per cent with the introduction of additional Gaussian components.}
In the second run, in order to measure the dust extinction (Sec. \ref{subsec:extinction}), we fit both \paa\ and \pab\ lines simultaneously in the same way as has been done for the \paa-only fits. In the fitting, the centroid velocities and line widths of \paa\ and \pab\ are tied to the same values while the line flux is allowed to vary freely. In both runs, we find that the emission line profiles, after quasar removal, can be fit adequately with only one or two Gaussian components. The stellar continuum emission of this source is very faint, and stellar \paa\ and \pab\ absorption features are not detected.{We further estimate the expected \paa\ and \pab\ EW based on \hb\ EW from Rupke et al. in prep. and PHOENIX synthetic stellar spectra \citep{phoenix}, and confirm that potential \paa\ and \pab\ absorption line flux are $\lesssim$1\% of the emission line flux.} Therefore, the emission line fitting is not compromised by fitting the continuum with polynomials instead of with stellar templates. The radial profile of the reconstructed PSF from our \qtdfit\ analysis is shown in Appendix \ref{sec:psf}, and we obtain a FWHM of $\sim$0.2\arcsec\ for the PSF.

\section{Results}
\label{sec:results}

\begin{figure*}
\includegraphics[width=1.0\textwidth]{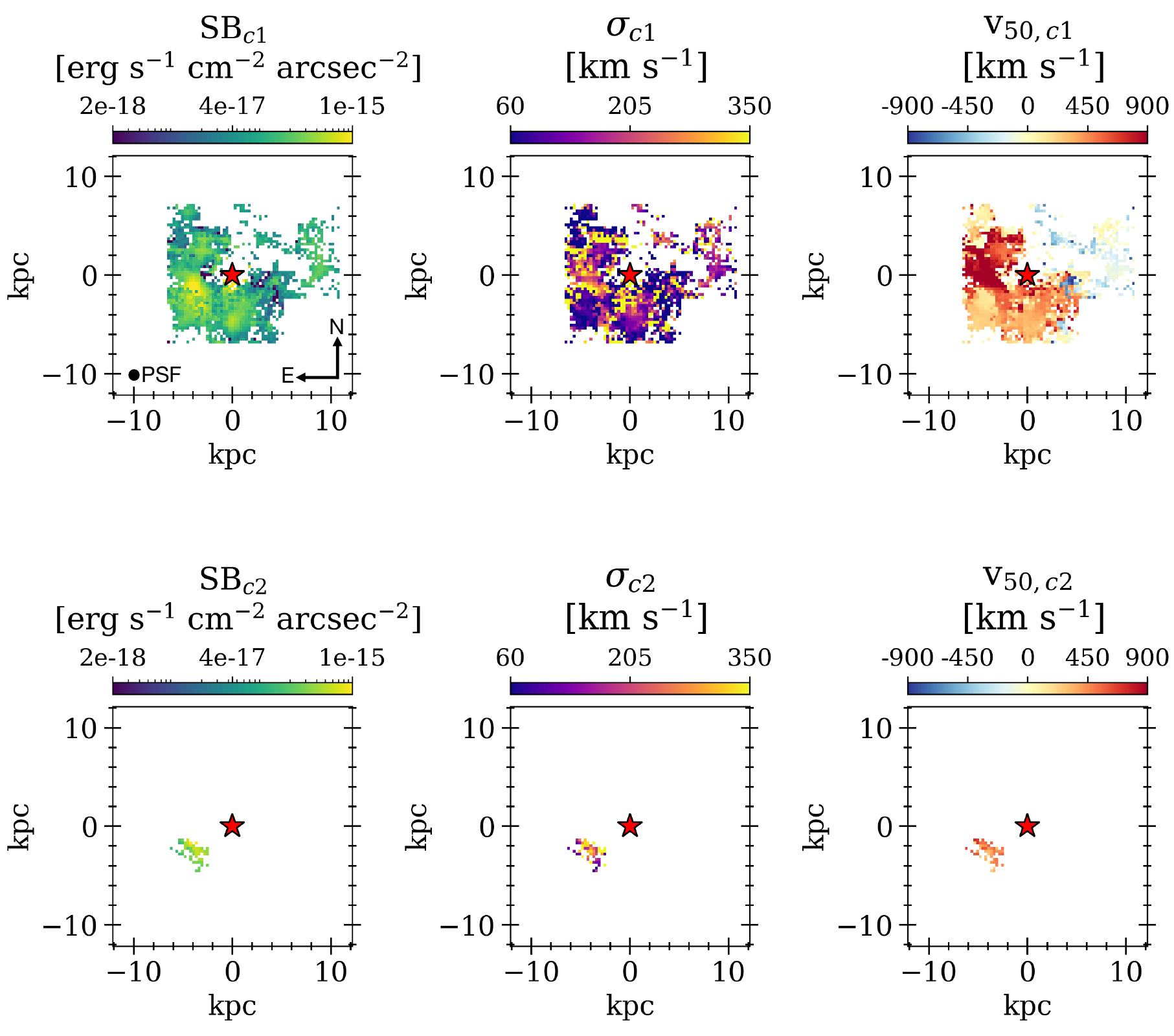}
\caption{Surface brightness, \vwu\ and $\sigma$ of the \paa-emitting gas in \ftm. The properties of the 1st Gaussian component $c1$ are shown in the top row, and those of the 2nd Gaussian component $c2$ are shown in the bottom row. The spatial scale is in unit of kpc. The location of the quasar is indicated by the red star.}
\label{fig:fig1}
\end{figure*}

\paa\ is the brightest emission line in the NIRSpec data cube, and a good tracer of the warm ionized gas in \ftm.  In the following sections, we examine in detail the properties of the extended \paa\ emission after quasar PSF removal.

\begin{figure*}
\includegraphics[width=1.0\textwidth]{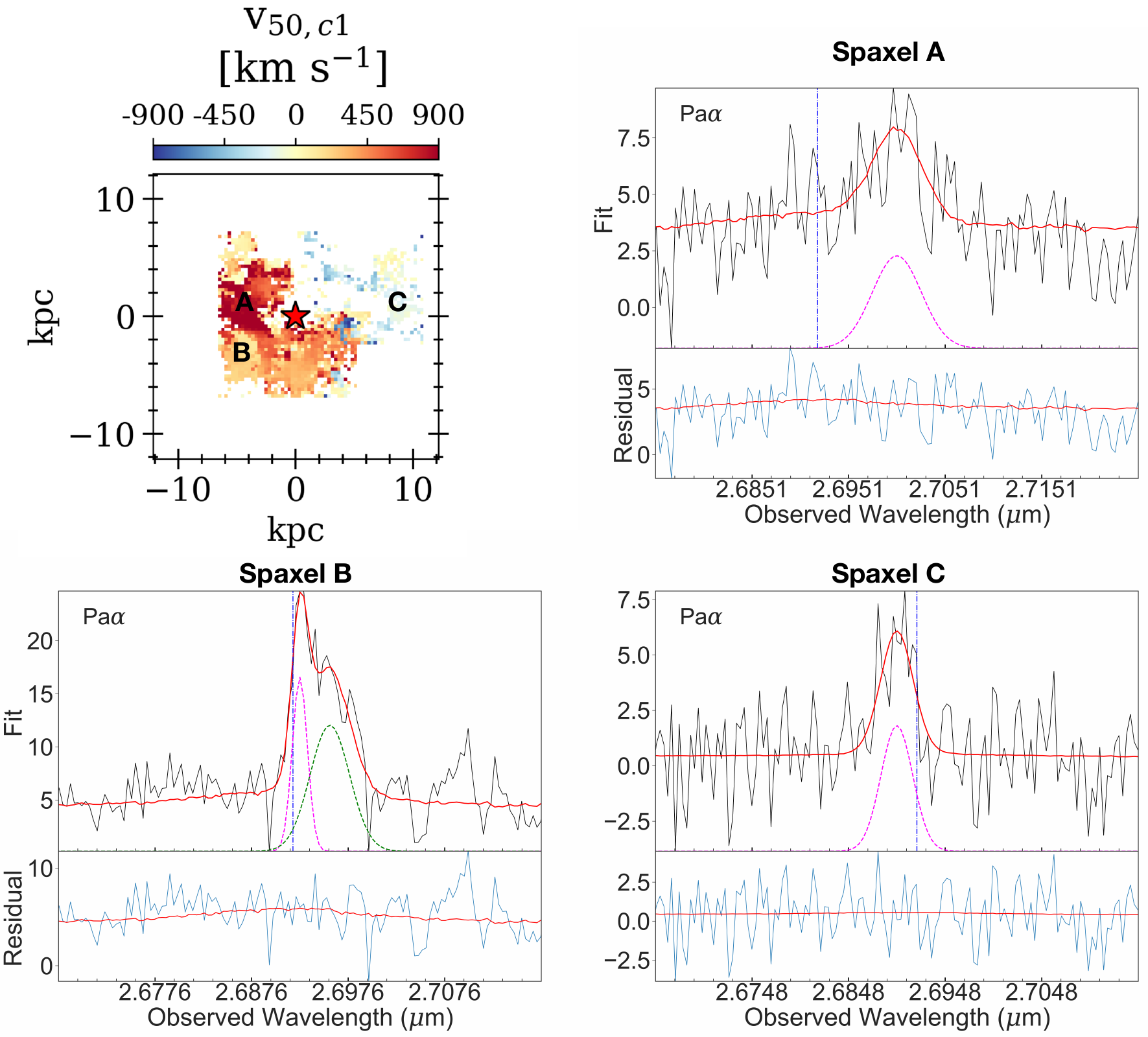}
\caption{Examples of \paa\ emission line profiles and best-fit models. \textbf{Top left:} The same \vwu\ map of the narrow Gaussian component as shown in the top right corner of Fig. \ref{fig:fig1}. \textbf{Top right and bottom panels:} Example spectra extracted from individual spaxels in the locations A, B and C shown in the \vwu\ map. In each of these three panels, the top plot shows the data in black (zoomed in to the \paa\ region), the overall best-fit model in red, and the individual Gaussian components of the fit in magenta and green dashed lines. The location of the line at systemic velocity is indicated by the blue dash-dotted line. The bottom plot shows the difference between the data and the best-fit host emission lines in blue, and the sum of the best-fit scaled quasar spectrum and host continuum model in red; see Sec.\ \ref{sec:analysis} for more detail). The y-axes in both plots are in the same arbitrary flux density units.}
\label{fig:fig2}
\end{figure*}

\begin{figure*}
\centering
\includegraphics[width=0.9\textwidth]{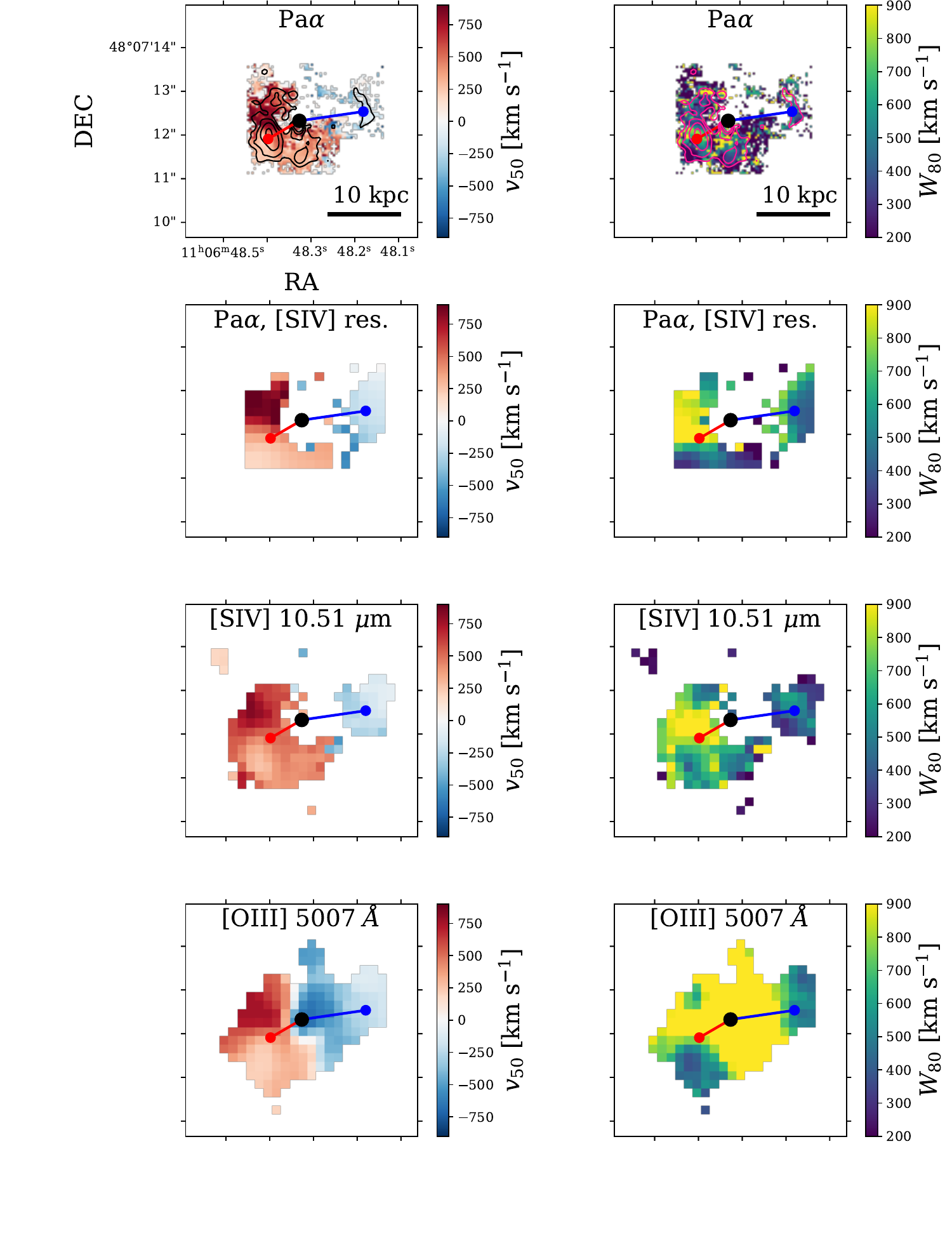}
\caption{From top to bottom, the velocity \vwu\ (left column) and line width \wba\ (right column) maps of \paa, \paa\ matched to \siv\ spatial resolution, \siv, and \oiii\ tracing the outflow in \ftm. The maps are plotted on the same spatial scale and color scale. As a guide for visual comparison, the location of the quasar is indicated by the black circle, and the \sivtext\ surface brightness peaks are plotted as red and blue circles, and they are connected to the quasar by red and blue lines. For the \paa\ maps in the first row, the black and pink contours are both the quasar-subtracted \paa\ surface brightness. The contour levels are 1.25$\times$10$^{-16}$, 2.50$\times$10$^{-16}$, 5.00$\times$10$^{-16}$ and 1.00$\times$10$^{-15}$ \fsb arcsec$^{-2}$. The \paa\ maps matched to the spatial resolution of \sivtext\ in the second row are measured from a data cube first convolved with a 2D Gaussian kernel to match the FWHM of the PSF of \sivtext\ and \oiiitext\ data and then binned to the spaxel size of the latter two.}
\label{fig:compare}
\end{figure*}

\subsection{\paa\ Morphology and Kinematics}
\label{subsec:morphology}

Fig. \ref{fig:fig1} shows the surface brightness and kinematic maps of \paa\ after the quasar emission has been subtracted by \qtdfit, and Fig. \ref{fig:fig2} shows examples of \paa\ emission line profiles from individual spaxels and their best-fit models in representative regions.

The \paa\ line-emitting gas outside the unresolved quasar is made up of two major components. One component is a redshifted nebula in the southeast with median radial velocity \vwu\ (velocity at the location that accumulates 50\% of the total flux, which equals the centroid velocity for a Gaussian profile) of $\sim$ 50 \kms\ to 1000 \kms\ and velocity dispersion $\sigma$ of $\sim$ 70 \kms\ to 690 \kms, with median values of $\sim$ 390 \kms\ and $\sim$ 180 \kms, respectively. The other component is a blueshifted nebula in the west with \vwu\ from $\sim-$870 \kms\ to 0 \kms\ and $\sigma$ of $\sim$ 70 \kms\ to 620 \kms, with median values of $\sim-$170 \kms\ and $\sim$ 180 \kms, respectively. The overall nebula extends to the east and south edges of the NIRSpec/IFS with a maximum spatial scale (in diameter) of $\sim$ 17 kpc in the east-west direction and $\sim$ 14 kpc in the north-south direction.

The median radial velocity \vwu\ of the redshifted nebula (examples: spaxels A and B in Fig. \ref{fig:fig2}) vary significantly, with the highest velocities seen to the east and lower velocities to the southeast and south. A single Gaussian component is adequate to describe all emission line profiles within this nebula except for a small region in the southeast, where a second Gaussian component is required. The spaxel B in Fig. \ref{fig:fig2} is an example of the \paa\ emission profile in this region. 
Meanwhile, the blueshifted nebula (example: spaxel C in Fig. \ref{fig:fig2}) is fainter and less extended when compared to the redshifted component. A single Gaussian component can describe the emission line profiles well in this nebula.

The bipolar morphology of the entire \paa\ nebula and the large velocities and velocity dispersions suggest that the \paa\ nebula is tracing a bipolar outflow, which is, as argued in Sec. \ref{sec:42} below, the same outflow discovered previously through \oiii\ with Gemini/GMOS in the rest-frame optical \citep[][]{Shen2023} and \siv\ with JWST MIRI/MRS \citep[][]{Rupke2023} in the rest-frame mid-infrared.

Moreover, the stellar component of the host galaxy seen in the KCWI data (Rupke et al. in prep.) and the warm molecular gas traced by H$_2$ (Diachenko et al. in prep.) both show completely different rotational fields with inverse velocity gradients (i.e., blueshifted to the east and redshifted to the west). These further support that the \paa\ nebula is tracing the rapid outflow.

\subsection{Comparison with \sivtext\ and \oiiitext\ Outflows}
\label{sec:42}

The outflow in \ftm\ has been detected via \siv\ \citep{Rupke2023}, \oiii\ \citep{Shen2023} previously. We compare the properties of the three outflow tracers (\paa, \sivtext, and \oiiitext) below.

The \sivtext\ was observed with JWST MIRI/MRS \citep{Rupke2023}, with a final spaxel size of 0.\arcsec2 and a spatial resolution of $\sim$0.\arcsec6–-0.\arcsec7.
The \oiiitext\ was observed with Gemini/GMOS IFS \citep{Shen2023}, with a spaxel size of 0.\arcsec2 and a spatial resolution of $\sim$0.\arcsec6.
In our comparison, we have coregistered the JWST/NIRSpec data (\paa), JWST/MIRI data (\sivtext) and Gemini/GMOS data (\oiiitext) to the same world coordinate system (WCS) using SDSS coordinates of the source as the references. This is done by aligning the flux centroids obtained from the 2D Gaussian fit to the narrow-band image integrated over each of the three data cubes. The wavelength ranges chosen for the integrations are 2.6--2.8 \mum\ for JWST/NIRSpec, 13.34–-15.57 \mum\ for JWST/MIRI, and 7300--7400 \AA\ for Gemini/GMOS.
The \vwu\ and \wba\ measurements for \sivtext\ and \oiiitext\ are those presented in the Fig. 2 of \citet{Rupke2023}. Here \wba\ is defined as the line width that comprises 80 percent of the total emission line flux.

As shown in Fig. \ref{fig:compare}, the morphology and kinematics of the \paa\ nebula are consistent with those of the \sivtext\ nebula (Fig. \ref{fig:compare}), except for the difference caused by their different spatial resolutions (FWHM of PSF: $\sim$0\arcsec.2 for \paa\ vs $\sim$0\arcsec.6 for \sivtext) and the smaller field of view for \paa\ than \sivtext. Due to the higher spatial resolution and thus less beam-smearing of \paa\ observation, the line width \wba\ map of \paa\ at the native spatial resolution (the first row in Fig. \ref{fig:compare}) exhibits more and sharper spatial variations and is thus less smooth than that of \sivtext. On the east of the object, there is a region with apparently higher \sivtext\ line width than \paa\ line width, which is mainly caused by the much heavier beam-smearing effect in the \sivtext\ map. This region with higher \sivtext\ line width surrounds the location where both \paa\ surface brightness and line width peak. If the \paa\ were instead observed with the spatial resolution of the \sivtext\ map, the beam smearing effect would increase the \paa\ line width for this region on the east to values comparable to those of \sivtext. To demonstrate this, we have convolved the \paa\ data cube to the spatial resolution of \sivtext\ data cube with a 2D Gaussian kernel and binned the spaxel to match that of the \sivtext\ cube (0.\arcsec2). The quasar PSF subtraction is then carried out with \qtdfit\ in the same way as has been done to the original \paa\ data cube, and the kinematic maps of the new \paa\ cube matched to the \sivtext\ resolution is shown in the second row of Fig. \ref{fig:compare}. The line width \wba\ of \paa\ and \sivtext\ on the east are now more consistent with each other once their spatial resolutions are matched, and so is the overall gas kinematics. In this region, the line width based on \sivtext\ are thus overestimated by up to a factor of $\sim$2, due to the heavier beam-smearing.

Similarly, we also find a good correspondence between the \paa\ nebula and \oiiitext\ nebula in the regions where both emission lines are detected (Fig. \ref{fig:compare}), except that there is a lack of \paa\ emission in the vicinity of the quasar where a band of high \oiiitext\ \wba\ runs from the north to the southwest, like the case for \sivtext\ as discussed in \citet{Rupke2023}. Given the similar spatial resolution between \sivtext\ and \oiiitext, the larger \oiiitext\ line width compared to that of \paa\ seen on the east is also caused by the heavier beam smearing effect in \oiiitext\ maps. The \paa\ maps matched to the spatial resolution of \sivtext\ are also in principle matched to that of \oiiitext, and these \paa\ maps are more consistent with \oiiitext\ maps than original \paa\ maps.   
However, the luminous broad \paa\ emission line from the quasar makes it extremely challenging to recover extended narrow \paa\ emission lines from the PSF subtraction in the regions adjacent to the quasar. It is thus possible that we are missing \paa\ emission lines tracing the outflow in the vicinity of the quasar that were captured by \oiiitext.

Overall, \paa, \sivtext, and \oiiitext\ observations altogether draw a consistent picture of the bipolar, $\sim$20 kpc outflow in \ftm.
However, without the higher spatial resolution of our \paa\ observation compared to those of \sivtext\ and \oiiitext, the line width would have been overestimated by up to a factor of $\sim$2 in certain regions of \ftm, and the outflow dynamics and energetics would also have been overestimated as a result. This emphasizes the importance of a good spatial resolution in recovering outflow properties in objects similar to \ftm. With a coarse spatial resolution (as the case for \sivtext\ and \oiiitext), we may easily obtain unreliable results.

\begin{figure}
\epsscale{1.1}
\plottwo{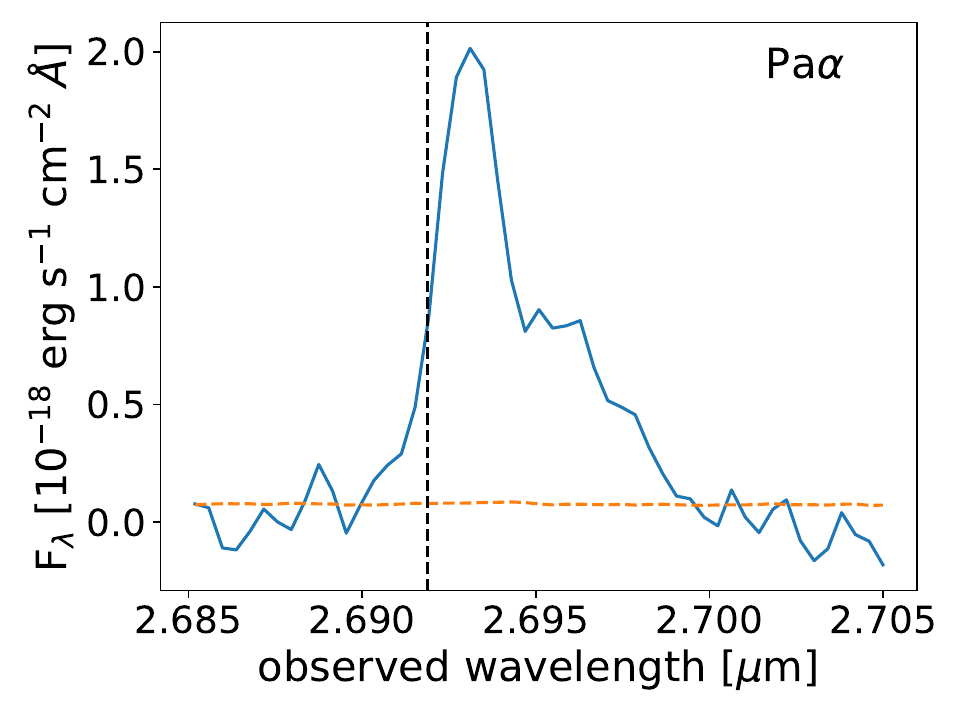}{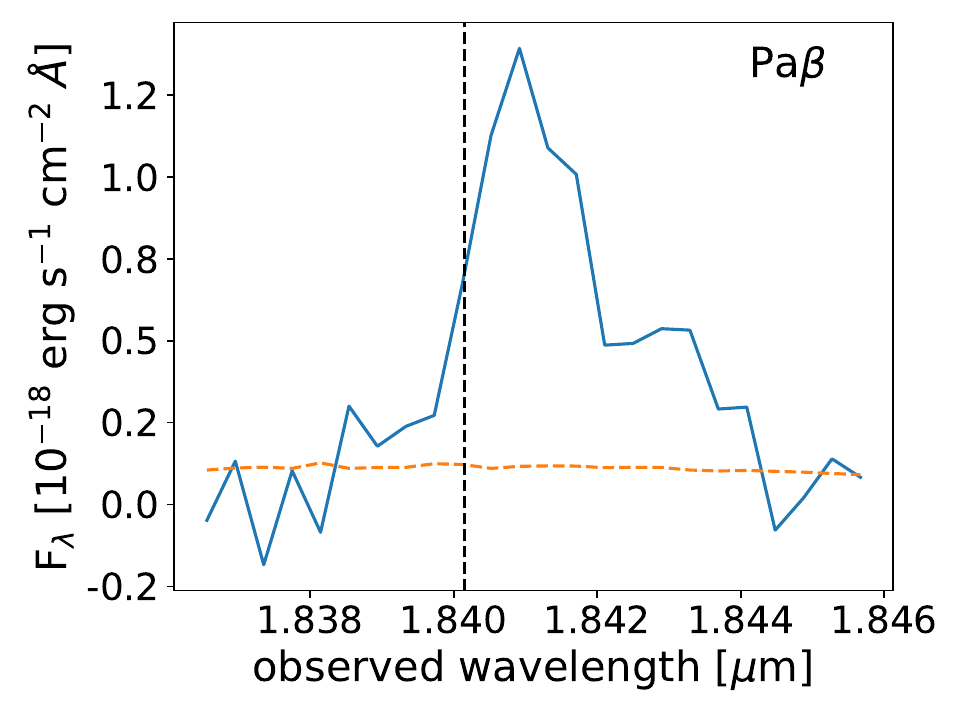}
\caption{Zoom-in spectra (blue) and spectral errors (orange) of \paa\ and \pab\ stacked over all spaxels with S/N(\pab) $>$ 2 from the quasar-subtracted cube. The location of the systemic velocity is indicated by the vertical dashed line.}
\label{fig:av}
\end{figure}

\subsection{Dust Extinction}
\label{subsec:extinction}

The dusty nature of red quasars like \ftm\ implies that the observed outflow is also likely dusty or attenuated by foreground dust. 
The dust extinction of the outflowing gas can be measured from the deviation of Hydrogen recombination line ratios from their intrinsic values. For our data, we adopt the two strongest emission lines \paa\ 1.87561\mum\ and \pab\ 1.28216\mum\ for this purpose, and assume an intrinsic \paa/\pab\ ratio of 2.049 \citep{HummerStorey87} under Case B condition and T$=$10$^{4}$ K. We then derive the \av\ from the \paa/\pab\ flux ratio adopting the extinction curve from \citet[][]{Chiar2006}, the same one as adopted in \citet{Rupke2023} when they estimated the \av.

The \paa/\pab\ ratios in many individual spaxels are highly uncertain, mainly due to the faintness of \pab\ emission after quasar PSF subtraction. 
To obtain a more robust estimate of the average dust extinction of the outflow nebula, we measure the \paa/\pab\ ratio from the spectrum stacked over all spaxels with S/N(\pab) $>$ 2 of the outflowing nebula, and obtain \paa/\pab$=$2.8$^{+0.2}_{-0.3}$ and thus \av$=$2.3$^{+0.7}_{-0.7}$, which is broadly consistent with the \av\ based on \sivtext/\oiiitext\ ratios ($\sim$0.3--2 with a median of $\sim$1.4) from \citet{Rupke2023}.

The spaxels with S/N(\pab) $>$ 2 are only found in the redshifted component of the outflow nebula. For the blueshifted component, we tried measuring the \pab\ flux from apertures with various sizes centered on the brightest spaxel of the component, but still obtained S/N(\pab) $\lesssim$2. This prevents a robust measurement of \paa/\pab\ ratio and thus dust extinction in this region.

\section{Discussion}
\label{sec:discussion}

Our \qtdfit\ analysis of the new NIRSpec data cube reveals the outflowing gas through \paa\ emission for the first time. Compared to previous studies based on ground-based observations of \oiiitext\ and JWST/MIRI observation of \sivtext, our new data provide a finer view (PSF $\sim$0.2\arcsec) of the fast ionized gas outflow on the spatial scale of $\sim$20 kpc. In this section, we evaluate the power source and impact of this outflow based on the new data.

\subsection{Energetics of the \paa\ Outflow}
\label{subsec:energetics}

We adopt \paa\ emission line (See Fig. \ref{fig:fig1}) to measure the mass of the outflowing gas. We substitute the \hb\ luminosity with \paa\ luminosity in the Eq. 1 from \citet{nesv11}, assuming case B conditions with an electron temperature $T \sim 10^4$ K \citep{Osterbrock2006}, which leads to: 
 \begin{eqnarray}
 M= 8.4\times 10^9 \left(\frac{L_{Pa\alpha}}{10^{43}\ {\rm erg\, s^{-1}}}\right)C_e\left(\frac{100\ {\rm cm}^{-3}}{n_e} \right) {\rm M_{\odot}}\hspace{10pt}
\label{eq:M_ionized}
\end{eqnarray}
where $L_{Pa\alpha}$ 
is the extinction-corrected luminosity of \paa. We adopt A$_{\rm V}=2.3$, the value measured from the stacked spectrum of the outflow nebula as described in Sec. \ref{subsec:extinction}, for the extinction correction. Note that this correction only boost the gas mass by $\sim$30\%, which reflects that \paa-based measurement is not very sensitive to dust extinction correction. The average electron density, $n_{e}$, is assumed to be 100 cm$^{-3}$, as adopted in previous studies of quasar-driven outflows at similar redshifts \citep[e.g.,][]{gree11,Liu13b}. The electron density clumping factor is defined as $C_e \equiv \langle n_e \rangle^2 / \langle n_e^2 \rangle $, which can be assumed to be of order unity on a cloud-by-cloud basis (i.e. each cloud has uniform density). In the end, we obtain an outflow mass of $\sim(5.7\pm{0.3})\times10^8$ \msun. {Note that the electron temperature of the outflowing gas may vary from the 10$^4$ K assumed above, and an increase/decrease of temperature by a factor of 2 leads to a decrease/increase of outflow mass by $\sim$2. However, this does not change our following results on outflow dynamics and energetics significantly.}

To calculate the mass, momentum and kinetic energy outflow rates, it is necessary to obtain the dynamical timescale of each parcel $i$ of outflowing gas, which is $t_{\rm dyn,i}$ $\approx$ ($R_{\rm deproj,i}/v_{\rm deproj,i}$) = ($R_i/v_i$), where $R_i$ represents the radial distance of the center of the gas parcel on the sky, and  $v_i = |v_{50}|+\sigma$ represents the outflow velocity of that same gas parcel. The outflow velocity is calculated in this form to account for the inclination correction needed for the outflow velocity in the 3D space. The inclusion of line width as the $\sigma$ term encapsulates both the outflow velocity not aligned with the line of sight and the turbulent motion of the outflowing gas. Such approach has been adopted in previous studies of quasar-driven outflows with NIRSpec IFS \citep[e.g.,][]{Vayner2023c}. Subsequently, the total mass, momentum and kinetic energy outflow rates can be obtained by integrating over all gas parcels, namely:

\begin{eqnarray}
\label{eq:Mdot}
\dot{M} & = & \Sigma~\dot{m_i} = \Sigma~m_i~(v_i/R_i).\\
\label{eq:pdot}
\dot{p} & = & \Sigma~\dot{m_i}~v_i,\\
\label{eq:Edot}
\dot{E} & = & \frac{1}{2}~\Sigma~\dot{m_i}(v_i)^2, 
\end{eqnarray}

The derived mass, momentum and kinetic energy outflow rates are $170\pm{40}$ \msunyr, $(1.3\pm{0.4})\times10^{36}$ dynes, and $(9.0\pm{3.7})\times10^{43}$ erg s$^{-1}$, respectively.
These outflow properties are also summarized in Table \ref{tab:energetics}.

The calculations of the outflow dynamics above assume that each parcel of the outflowing gas has traveled from the center of the galaxy to its current location on the sky plane and has its own dynamical time scale. This may underestimate the distance each outflowing gas parcel has traveled, mainly because the distance that the outflow has traveled along the line-of-sight is not accounted for. As a result, the total outflow rates may be overestimated. 

To estimate the systematic uncertainties associated with the outflow dynamics and energetics caused by the assumption made above, we can consider an alternative scenario where all the outflowing gas parcels have traveled the same distance from the galaxy center and share the same dynamical timescale. Here the distance that the outflow has traveled is the maximum radius of the observed \paa\ nebula. In this scenario, on the contrary, the distance that the majority of the outflowing gas parcels have traveled are likely overestimated, and the total outflow rates are thus underestimated.  The corresponding mass, momentum, and kinetic energy outflow rates can instead be estimated following:

\begin{eqnarray}
\label{eq:Mdot2}
\dot{M} & = & M_{out}v_{out}/R_{out} = (\Sigma~m_i) v_{out}/R_{out}.\\
\label{eq:pdot2}
\dot{p} & = & \dot{M}v_{out},\\
\label{eq:Edot2}
\dot{E} & = & \frac{1}{2}~\dot{M}v_{out}^2, 
\end{eqnarray}

Here M$_{out}$ is the total outflow gas mass derived from eq. \ref{eq:M_ionized}. The outflow velocity $v_{out}$ is defined as the maximum value of $|v_{50}|+\sigma$, which is 1120$\pm{80}$ \kms. $R_{out}$ is the maximum radial distance of the outflow (i.e., maximum radius of the \paa\ nebula), which is $\sim$12 kpc. These lead to mass, momentum and kinetic energy outflow rates of $60\pm{10}$ \msunyr, $(3.8\pm{0.8})\times10^{35}$ dynes, and  $(2.2\pm{0.6})\times10^{43}$ erg s$^{-1}$, respectively, as listed in Table \ref{tab:energetics}. 

{Overall, the IFU data only provide outflow distance in the sky plane, but not along the line-of-sight. To account for the uncertainties of outflow rates caused by the distance uncertainties, we have adopted two approaches to calculate the total time-averaged outflow rates: In eq. \ref{eq:Mdot}, \ref{eq:pdot} and \ref{eq:Edot}, we use the distance of each spaxel from the AGN; In e.q. \ref{eq:Mdot2}, \ref{eq:pdot2} and \ref{eq:Edot2}, we adopt a constant distance for the entire outflow, which effectively assumes a single radius, spherically-symmetric outflow. The differences in the outflow rates derived from the two scenarios above thus reflect the systematic uncertainties associated with the outflow rates, which is a factor of $\sim$3--4.}

\begin{deluxetable*}{cccc ccc cc}[!htb]
\tablecolumns{9}
\tabletypesize{\scriptsize}
\tablecaption{Properties of the Outflow\label{tab:energetics}}
\tablehead{ \colhead{$v_{max}$ } &
  \colhead{$M_{out}$}  & \colhead{R$_{out}$}    & \colhead{$\dot{M}_{out}$} & \colhead{$\dot{p}_{out}$ }  & \colhead{$\dot{E}_{out}$} & \colhead{$\dot{p}_{out}$/(L$_{qso}/c$)} &\colhead{$\dot{E}_{out}$/L$_
  {qso}$} &  \colhead{Eq.}  \\
\colhead{[km s$^{-1}$]} & \colhead{[\msun]} & \colhead{[kpc]} & \colhead{[\msunyr]} & \colhead{[dynes]} & \colhead{[erg s$^{-1}$]} & &  & \\
 & $\times(\frac{n_e}{100\ \mbox{cm}^{-2}})^{-1}$ &  & $\times(\frac{n_e}{100\ \mbox{cm}^{-2}})^{-1}$ & $\times(\frac{n_e}{100\ \mbox{cm}^{-2}})^{-1}$ & $\times(\frac{n_e}{100\ \mbox{cm}^{-2}})^{-1}$ & $\times(\frac{n_e}{100\ \mbox{cm}^{-2}})^{-1}$ &  $\times(\frac{n_e}{100\ \mbox{cm}^{-2}})^{-1}$ & \\
\colhead{(1)} & \colhead{(2)} & \colhead{(3)} & \colhead{(4)} & 
\colhead{(5)} & 
\colhead{(6)} &  
\colhead{(7)} & \colhead{(8)} & \colhead{(9)} 
}
\startdata  
\multirow{2}{*}{$1120\pm{80}$} &
 \multirow{2}{*}{$(5.7\pm{0.3})\times10^8$} & \multirow{2}{*}{12} & $170\pm{40}$ & $(1.3\pm{0.4})\times10^{36}$ & $(9.0\pm{3.7})\times10^{43}$ & 0.60$\pm{0.18}$ &  0.14$\pm{0.06}$\% &(2)-(4)  \\
 & & & $60\pm{10}$ & $(3.8\pm{0.8})\times10^{35}$ & $(2.2\pm{0.6})\times10^{43}$ & 0.18$\pm{0.04}$ &  0.03$\pm{0.01}$\% & (5)-(7) 
\enddata
\tablecomments{Outflow properties based on the \paa\ emission line. From left to right, the columns are (1) outflow velocity defined as the 90 percentile value of $|v_{50}|+\sigma$ for all spaxels ranking from low to high (i.e., the maximum value excluding the 10\% spaxels with the highest $|v_{50}|+\sigma$); (2) outflow mass;  (3) maximum radial distance of the outflow; (4) mass outflow rate; (5) momentum outflow rate; (6) kinetic energy outflow rate; (7) momentum outflow rate to quasar photon momentum flux; (8) kinetic energy outflow rate to the quasar bolometric luminosity; (9) equation references adopted to derive column (4)--(8). The first row lists spatially integrated values adopting Eq. \ref{eq:Mdot}, \ref{eq:pdot} and \ref{eq:Edot}, and the second row lists values assuming a single dynamical time scale adopting Eq. \ref{eq:Mdot2}, \ref{eq:pdot2} and \ref{eq:Edot2} for the calculations. See Sec. \ref{subsec:energetics} for more details. }
\end{deluxetable*}

\subsection{Power Source of the Outflow}
\label{sec:52}
In the following discussion, we evaluate the energy source of the observed outflow. For the outflow dynamics and energetics, we adopt the two sets of values listed in Table \ref{tab:energetics}, derived from the two scenarios stated in Sec. \ref{subsec:energetics}, as their lower and upper limits. 

Given the quasar bolometric luminosity of $L_{\rm qso} \sim 6.5 \times 10^{46}$ erg s$^{-1}$ based on rest-frame 12 \mum\ luminosity \citep{Shen2023}, the measured outflow momentum rate $\dot{p}$ ($\sim(0.3-1.7)\times10^{36}$ dynes) are $\sim$0.14--0.78$\times$ the radiative force provided by the quasar, $\dot{p}_{qso} = L_{\rm qso}/c$ $\approx$ 2.2 $\times$ 10$^{36}$ dynes. The quasar is in principle capable of driving this outflow via radiation pressure. 
The ratio of the kinetic energy outflow rate ((0.16--1.27)$\times10^{44}$ erg s$^{-1}$) to the quasar bolometric luminosity, $\dot{E}/L_{\rm qso} \simeq 0.02-0.20\%$, is within the typical range seen in type 1 quasar with similar luminosity \citep[e.g.][]{Rupk17,Har2018}. The quasar is in principle powerful enough to drive the observed outflow easily.

On the other hand, it is also important to investigate the possible role of star formation activity in driving the outflow.
Adopting the best-fit result from the spectral energy distribution (SED) modeling of \ftm\ (Sankar et al. in prep.), the star formation rate (SFR) of the object is $\sim$132 \msunyr\ based on the 8-10000 \mum\ infrared luminosity. Therefore, the mass outflow rate ($\sim$50--210 \msunyr) is likely comparable to the star formation rate in \ftm\ based on the SED fitting. The implied mass-loading factor, $\dot{M}$/{SFR}, is $\sim$0.4--1.5, which is higher than or comparable to the mass-loading factors for starburst-driven outflows that are in general less than unity \citep[e.g.][]{Arribas2014}. The energy output rate from a starburst with SFR $=$132 \msun\ yr$^{-1}$ is $\dot{E_*}$ $\approx$ 7 $\times$ $10^{41}$ (SFR/\msun\ yr$^{-1}$) erg s$^{-1}$ $\approx$ 9 $\times$ 10$^{43}$ erg s$^{-1}$ \citep[e.g.][]{Vei2005}. Assuming that the energy output is coupled to the outflow at 100\% efficiency (which is unlikely in reality), the energy injected to the outflow is $\sim$ 18--140\% of the measured kinetic power $\dot{E}$ in the outflow. Therefore, it may be challenging for stellar processes to drive the outflow alone.
Overall, the quasar can easily drive the observed outflow. However, we cannot formally rule out the possibility that the star formation activity also contributes to the launching of the outflow. {Note that the outflow rates are proportional to $n_e^{-1}$ as shown in eq. \ref{eq:M_ionized}, and the values listed in Table \ref{tab:energetics} are all adopting n$_e=$ 100 cm$^{-3}$. An increase from $n_e=$ 100 cm$^{-3}$ to $n_e=$ 1000 cm$^{-3}$ \citep[The typical maximum value reported for quasar-driven outflows; e.g.,][]{Harrison2018} instead makes the mass, momentum, and kinetic energy outflow rates 10\%\ of the values reported above. In such case, it becomes more plausible that the star formation activity may play an important role in driving the outflow.}

\begin{figure*}
\includegraphics[width=\textwidth]{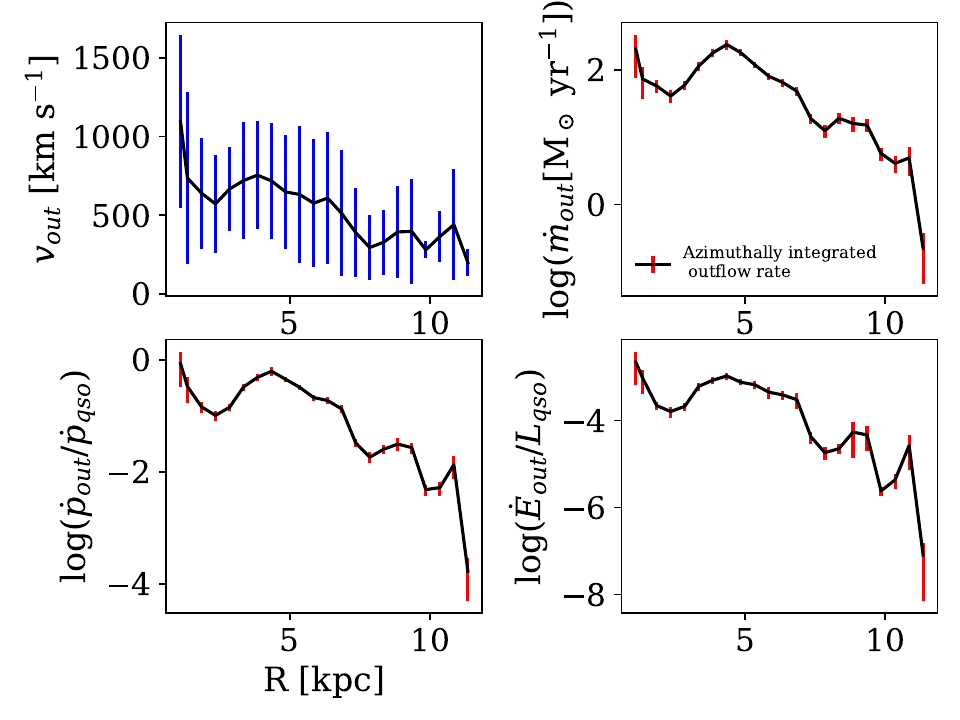}
\caption{Radial profiles of the outflow velocity (upper left), mass outflow rate (upper right), the ratio of momentum outflow rate to radiative force from the quasar (bottom left), and the ratio of kinetic energy outflow rate to quasar bolometric luminosity (bottom right). {Here the outflow rates are all local instantaneous outflow rates, which are averaged over the time for the outflow to travel the physical distance of a spaxel (eq. \ref{eq:mdot_r}, \ref{eq:pdot_r} and \ref{eq:Edot_r}) and are local properties.} In the upper left panel, the data points and blue vertical bars are the median values and standard deviations of outflow velocities within individual radial bins. In the other three panels, the data points are the azimuthally integrated mass, momentum and kinetic energy outflow rates in individual radial bins, respectively. The measurement uncertainties are shown as red error bars.}
\label{fig:radial}
\end{figure*}

Assuming that the outflow is driven by the quasar, we further investigate how the outflow is launched and propagates within the system, by investigating the radial profiles of the velocity and local instantaneous mass, momentum and kinetic energy outflow rates (Fig. \ref{fig:radial}). For the radial profile of velocity, we measure the median value and standard deviation of each radial bin, with a starting radius of 0.6 kpc and bin size of 0.5 kpc. For the radial profiles of outflow rates, we first calculate the values of individual spaxels following:

\begin{eqnarray}
\label{eq:mdot_r}
\dot{m}_{i,inst.} & = &m_iv_i/\Delta R.\\
\label{eq:pdot_r}
\dot{p}_{i,inst.} & = & ~\dot{m}_{i,inst.}v_i(r),\\
\label{eq:Edot_r}
\dot{E}_{i,inst.} & = & \frac{1}{2}~\dot{m}_{i,inst.}(v_i)^2, 
\end{eqnarray}
Here $\Delta$R is the physical distance of an individual spaxel. All other variables are the same as those in eq. \ref{eq:Mdot}, \ref{eq:pdot} and \ref{eq:Edot}. 
The radial profiles are then calculated as azimuthal integrations of these instantaneous mass, momentum, and kinetic energy rates in individual radial bins.
As shown in Fig. \ref{fig:radial}, the outflow velocity and instantaneous mass, momentum, and kinetic energy outflow rates decrease with increasing radius in general. There are local maxima of instantaneous outflow rates at $\sim$4 kpc, which are comparable to the global maxima of the outflow.

In the following, we focus on two major driving mechanisms commonly considered in the literature \citep[e.g.,][]{King2015ARAA}, which are the radiation pressure and the ``blast-wave'' model, respectively.

For radiation pressure-driven outflows, the ratio of momentum outflow rate to quasar photon momentum flux $\dot{p}_{outflow}$/$\dot{p}_{qso}$ is usually $\lesssim$1 but can reach as high as $\sim$2 on kpc scales if the infrared photons are trapped in very high column density environments \citep{Thompson2015,Bieri2017,Costa2018}. In this scenario, at any given time, the $\dot{p}_{outflow}$/$\dot{p}_{qso}$ tends to peak at small radius and decrease outwards \citep[e.g., see Sec. 3.1 and Fig. 4 of][]{Thompson2015}. 

Alternatively, in the ``blast-wave'' model, quasar-driven winds, from either accretion disks or jets, can shock the immediate interstellar medium (ISM) and sweep up material while travelling outwards, which results in the observed galactic-scale outflow. Depending on various factors including nuclear density, wind/jet velocity, and shock velocity, the shock may expand adiabatically and cool rapidly and efficiently, or cool inefficiently and expand in an energy-conserving manner \citep[][]{fauc12b,Costa14}.
Specifically, the energy-conserving shock preserves a significant fraction of the initial energy, leading to a $\dot{p}_{outflow}$/$\dot{p}_{qso}$ ratio of $\sim$7--10 for kpc-scale outflows. At any given time, the radial profile of $\dot{p}_{outflow}$/$\dot{p}_{qso}$ tends to increase outwards on spatial scale $\lesssim$ 10 kpc \citep[e.g., see Sec. 3.3, 3.4 and Fig. 4 of][]{fauc12b}, in contrast with a radiation-driven outflow.

Given a total $\dot{p}_{outflow}$/$\dot{p}_{qso}$ of $\sim$0.64 and a radial profile of $\dot{p}_{outflow}$/$\dot{p}_{qso}$ peaking at small radius and generally decreasing outward, the outflow in \ftm\ is more consistent with a radiatively-driven galactic outflow.

\subsection{Impact of the Quasar and Outflow on the Host Galaxy}
\label{subsec:feedback}

The mass outflow rate in \ftm\ is $\sim$ 0.4--1.5$\times$ SFR. The outflow is thus capable of expelling a significant amount of gas outwards, and thus reducing the amount of gas available for star formation substantially.

The kinetic energy outflow rate is $\sim$ 0.02--0.20\% of the quasar bolometric luminosity, which is comparable to the minimum value required for negative quasar feedback in simulations \citep[e.g.,][]{Choi2012,Hopkins2012}. The outflow may help regulate the star formation activity within the host galaxies to some extent.

One important caveat for the analysis above is that the mass, momentum and kinetic energy outflow rates adopted are likely lower limits for the total outflow rates. In the calculation, we have only considered the extended, spatially resolved outflow, and ignored the spatially unresolved, nuclear outflow in the vicinity of the quasar. The luminous, broad \paa\ emission line from the broad line region of the quasar makes it impossible to recover the spatially unresolved outflow reliably. However, the energetics of the unresolved outflow is expected to be significant in such dusty quasars with powerful outflows, as is the case for another quasar in ``Q-3D'' program \citep{Vayner2023c}, SDSS J1652$+$1728. Additionally, in many similar objects, the quasar-driven outflows are usually multi-phase \citep[i.e., ionized, neutral and molecular phases; e.g.,][]{Veilleux20}. The higher outflow rates, as expected, will further strengthen our main conclusions, including that the outflow is driven by the quasar, and that the outflow may provide negative feedback to the galaxy. {As emphasized in Sec. \ref{sec:52} the outflow rates are proportional to $n_e^{-1}$ as shown in eq. \ref{eq:M_ionized}. The $n_e=$ 100 adopted in our calculations may be underestimated. The derived outflow rates and thus the impact of the outflows may be overestimated. }

Furthermore, as shown in Fig. \ref{fig:radial}, the local instantaneous mass, momentum and kinetic energy outflow rates are in general decreasing radially, suggesting that the outflow may have a stronger overall impact on the inner part of the galaxy than in the outer part. In addition to the general decreasing trends, there are local maxima of instantaneous outflow rates at $\sim$4 kpc, which are comparable to the global maxima. This suggests that the outflow still carrys a significant amount of momentum and energy that may be deposited to the system after travelling several kpc in our object.

\section{Summary}
\label{sec:conclusions}

In this paper, we examine the JWST/NIRSpec IFS data of \ftm, a red quasar at $z=0.4352$ with a known powerful outflow. Our main results are summarized as follows.

\begin{itemize}

\item
We detect the \paa\ nebula tracing the fast outflow on galactic-scale, after quasar PSF removal.
The outflow shows a bipolar morphology, with the redshifted component to the south and east and the blueshifted component to the northwest of the quasar. The redshifted component shows a maximum (median) velocity \vwu\ of $\sim$ 1000 (390) \kms\ and maximum (median) velocity dispersion of $\sim$ 690 (180) \kms. The blueshifted component shows an absolute maximum (median) velocity $|$\vwu$|$ of $\sim$ 870 (170) \kms\ and maximum (median) velocity dispersion of $\sim$ 620 (180) \kms. The outflow extends up to $\sim$17 kpc in the east-west direction and $\sim$14 kpc in the north-south direction.

\item 
The outflow detected in \paa\ is in good agreement with those detected in \siv\ \citep{Rupke2023} and \oiii\ \citep{Shen2023} in regions where all three emission lines are detected. The overall morphology and kinematics of the outflow are consistent across all three tracers, except that the outflowing gas detected in \oiiitext\ in the vicinity of the quasar is undetected in both \paa\ and \sivtext. For \paa, such non-detection is likely due to the challenge in removing the luminous quasar PSF reliably. Additionally, the higher spatial resolution (less beam-smearing) of \paa\ observation allows for more reliable measurements of outflow kinematics, compared to the \sivtext\ observation. 

\item
We estimate the dust extinction of the outflow by comparing the observed \paa/\pab\ ratio from the stacked spectrum to the intrinsic value, and obtain an \av\ of $2.3^{+0.7}_{-0.7}$, which is broadly consistent with the \av\ measured from \sivtext/\oiiitext\ ($\sim$0.3--2 with a median of $\sim$1.4) from \citet{Rupke2023}.  
\\

\item 
Adopting an electron density of 100 cm$^{-2}$, we obtain mass, momentum, and kinetic energy outflow rates of 50--210 \msunyr, (0.3--1.7)$\times10^{36}$ dynes, and (0.16--1.27)$\times10^{44}$ erg s$^{-1}$, for the spatially resolved outflow, respectively. The momentum outflow rate is $\sim$ 14--78\% of the quasar photon momentum flux, and the kinetic energy outflow rate is $\sim$ 0.02--0.20\% of the quasar bolometric luminosity, suggesting that the quasar is powerful enough to drive the outflow. 
The local instantaneous outflow rates in general decrease towards larger radial distance, which is in principle consistent with a radiatively driven galactic outflow and hints that the overall impact of the outflow may be stronger in the inner part of the system than the outer part. Additionally, there are local maxima of
instantaneous outflow rates at $\sim$4 kpc, which are comparable to the global maxima of the outflow.
For the entire outflow, the mass-loading factor, $\dot{M}$/{SFR}, is $\sim$ 0.4--1.5. The output energy rate from star formation activity is $\sim$ 18--140\% of the measured kinetic energy outflow rate. While it is challenging for the stellar processes to drive the outflow alone, we cannot formally rule out the possibility that they may help launch the outflow.

\item 
The mass outflow rate is comparable to the star formation rate, suggesting that the outflow is capable of driving gas outwards and thus reducing the amount of gas available for star formation.
The ratio of kinetic energy outflow rate to quasar bolometric luminosity is comparable to the minimum requirement ($\sim$0.1\%) for negative feedback as predicted by simulations. The outflow may help regulate the star formation activity within the system to some extent.

\end{itemize}


\begin{acknowledgments}

W.L., S.V., A.V., D.S.N.R., and N.L.Z. were supported in part by NASA through STScI grant JWST-ERS-01335. D.W. and C.B. acknowledge support through an Emmy Noether Grant of the German Research Foundation, a stipend by the Daimler and Benz Foundation and a Verbundforschung grant by the German Space Agency. J.B.-B. acknowledges support from the grant IA- 101522 (DGAPA-PAPIIT, UNAM) and funding from the CONACYT grant CF19-39578.

\end{acknowledgments}

\appendix
\section{Reconstructed PSF}\label{sec:psf}

\begin{figure}[!b]
    \centering
    \includegraphics[width=0.5\linewidth]{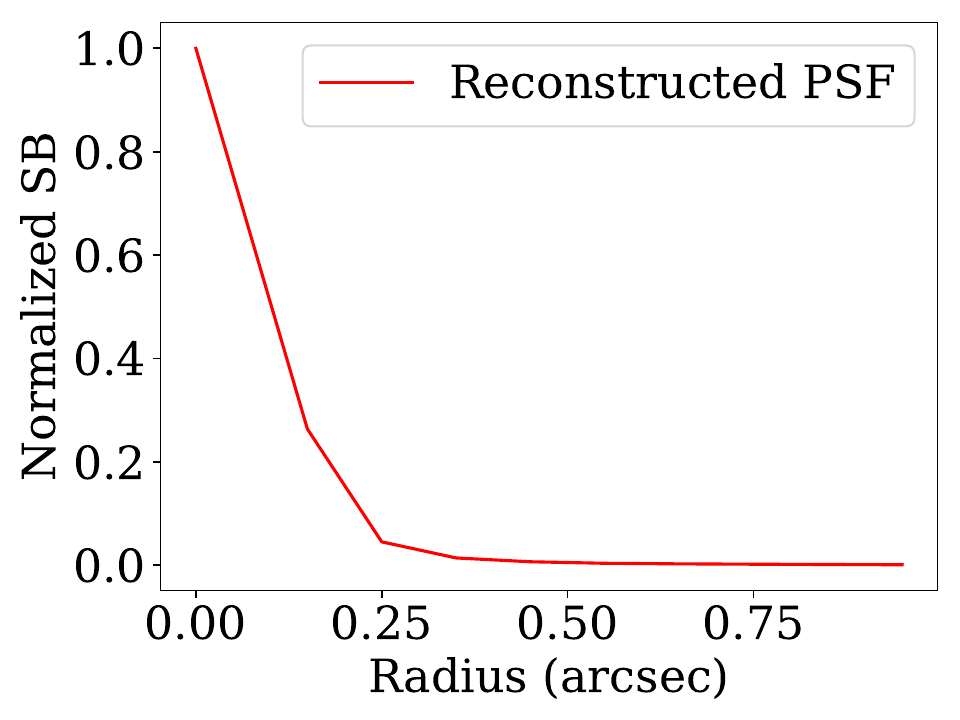}
    \caption{The normalized, azimuthally-averaged surface brightness (SB) radial profile of the reconstructed PSF at 2.6--2.8 \mum.}
    \label{fig:psf}
\end{figure}

The normalized, azimuthally-averaged radial profile of the reconstructed PSF at 2.6--2.8 \mum\ from \qtdfit\ is shown in Fig. \ref{fig:psf}.

%

\vspace{5mm}
\facilities{JWST(NIRSpec), Gemini(GMOS) }


\software{astropy \citep{Ast2013, Ast2018},  \qtdfit\ \citep{q3dfit}}





\bibliography{f2m}{}
\bibliographystyle{aasjournal}



\end{document}